\documentstyle [12pt]{article}
\makeatletter \oddsidemargin  0in \evensidemargin 0in
\textwidth16cm \RequirePackage[dvips]{graphicx} \textheight 20.5cm
\newcommand{\singlespacing}{\let\CS=\@currsize\renewcommand{\baselinestretch}{1.5}\tiny\CS}
\makeatletter \oddsidemargin  -.1in \evensidemargin -.1in
\newcommand{\doublespacing}{\let\CS=\@currsize\renewcommand{\baselinestretch}{1.35}\tiny\CS}

\def\@citex[#1]#2{\if@filesw\immediate\write\@auxout{\string\citation{#2}}\fi
  \def\@citea{}\@cite{\@for\@citeb:=#2\do
    {\@citea\def\@citea{,\linebreak[0]\hskip0pt plus .2em}%
      \@ifundefined{b@\@citeb}%
    {{\bf ?}\@warning{Citation `\@citeb' on page \thepage\space undefined}}%
      \hbox{\csname b@\@citeb\endcsname}}}{#1}}

\newtheorem{rule-def}[theorem]{Rule}
\begin{document}
\title{\bf Witness for edge states and its characteristics}
\author{Nirman Ganguly$^{1}$\thanks{nirmanganguly@rediffmail.com}, Satyabrata Adhikari$^{2}$\\
$^1$ Heritage Institute of Technology, Kolkata-107, West Bengal,India\\
$^2$ S.N.Bose National Centre for Basic Sciences,Salt lake,Kolkata
700098,\\West Bengal, India \\}

\date{}
\maketitle{}
\begin{abstract}
Edge states lying at the edge of PPT entangled states have a very
intriguing existence and their detection is equally interesting.
We present here a new witness for detection of edge states. We
then compare between our proposed witness operator and the witness
operator proposed in [Physical Review A, 62, 052310 (2000)] in
terms of the efficiency in the detection of PPT entangled states.
In this regard we show that this operator is finer than the
Lewenstein et.al. operator in some restriction. We also discuss
about its experimental realization via Gell-Mann matrices.
\end{abstract}

\section{Introduction }
Quantum entanglement \cite{1,2} is one of the most amazing
features of quantum formalism. Its spooky features makes it an
enigma drawing attention from scientists worldwide over the years.
The development of its knowledge theoretically and experimentally
made possible a number of practical applications including quantum
computation \cite{3} and quantum teleportation \cite{4}.\\
The significance of entanglement in quantum information theory
makes its distinction from separable states all the more
important. For low dimensional (2 $\otimes$ 2 and 2 $\otimes$ 3)
states there exist simple necessary and sufficient conditions for
separability \cite{5,6} which is based on the fact that separable
states have a positive partial transpose (PPT). For higher
dimensional systems all states with negative partial transpose
(NPT) are entangled but there are entangled states which have a
positive partial transposition \cite{7,8}. Thus the separability
problem can be framed as finding whether states with positive
partial transposition are entangled. Of specific importance in
this context are the so called edge states \cite{9} which lies at
the boundary of PPT and NPT states. An interesting character that
an edge state shows is extreme violation of the range criterion
\cite{7} which states that there exists no product vector
$|e,f\rangle$ belonging to the range of the edge state $\varrho$
such that $|e,f^{*}\rangle$(conjugation is done with respect to
the second system) belongs to the range of $\varrho^{T_{B}}$.
Since the edge states are PPT entangled states so partial
transposition method fails to detect them and also it is very
difficult to identify the edge states by range criterion. So it
becomes necessary to find an alternate method to detect the
edge states.\\
A very general method to distinguish between entangled and
separable states is through witness operators \cite{6,10}. Witness
operator is the outcome of the celebrated Hahn-Banach theorem in
functional analysis. It is a hermitian operator, thus observable
with at least one negative eigenvalue. The witness operators act
as a hyperplane separating separable states from entangled ones.
They can be divided into two classes: Decomposable witness (DW)
operators and Non-decomposable witness (NDW) operators. DW can
detect only NPT states while NDW detects not only NPT states but
also PPT entangled states. Terhal first introduced a family of
indecomposable positive linear maps based on entangled quantum
states \cite{10} using the notion of unextendible product basis.
Thereafter Lewenstein et. al. extensively worked on indecomposable
witnesses and provided an algorithm to optimize them \cite{12}.
These operators are also of prime importance because they can be
used in an experimental set up to detect inseparability. An
experimental realization of a geometric entanglement witness in
terms of Gell-Mann matrices and spin-1 operators was studied in
\cite{13}. This makes witness
operators all the more significant from a very pragmatic sense.\\
In this paper together with the proposition of a new witness for
edge states, we show that our proposed witness operator is finer
than the witness introduced in \cite{12} in some cases. We also
provide an insight as to how an experimental realization can be done of our proposed witness. \\
Our paper is organized as follows: In section 2 we review certain
related definitions and terms. In section 3 we revisit the
non-decomposable witness operator and find the condition for which
it is finer. In section 4 we give the construction of the witness,
its extension to multipartite edge states and discuss its
experimental realization. In section 5 we compare our proposed
witness with that in \cite{9,12}. In section 6 we provide explicit
examples. Lastly we end with conclusions.
\section{Prerequisites: A few definitions and results}
\textbf{Definition-1:} The kernel of a given density matrix $\rho
\in B(H_{A}\otimes H_{B})$ is defined as the set of all
eigenvectors corresponding to the zero eigenvalue in the Hilbert
space $H_{A}$. Mathematically, $ker(\rho)=\{|x\rangle\in
H_{A}:\rho|x\rangle=0\}$.\\
\textbf{Definition-2:} A PPT entangled state $\delta$ is called an
edge state if for any $\varepsilon>0$ and any product vector
$|e,f\rangle$, $\delta^{'}=\delta-\varepsilon|e,f\rangle\langle
e,f|$ is not a PPT state.\\
\textbf{Definition-3:} A hermitian operator $W$ is said to be an
entanglement witness if the expectation value of $W$ is negative
for entangled state whereas it is non-negative for any separable
state. Mathematically, it can be formulated as
\begin{eqnarray}
&&(i)~ Tr(W\sigma)\geq 0~~~  \forall~~  \textrm{separable state}~
\sigma~~ and \nonumber\\&& (ii)~Tr(W\rho)<0 ,~~ \textrm{for at
least one entangled state}~\rho. \label{def.3}
\end{eqnarray}
\textbf{Definition-4:} A witness operator is said to be
decomposable if it can be expressed in the form
\begin{eqnarray}
D= P+Q^{T_{B}}\label{decomwit.}
\end{eqnarray}
where $P$ and $Q$ are positive semi-definite operators.\\
Non-decomposable operators are those which cannot be written as in
(\ref{decomwit.}).\\
\textbf{Definition-5:} Given two entanglement witnesses $W_{1}$
and $W_{2}$, a witness $W_{1}$ is said to be finer than another
witness $W_{2}$ if $D_{W_{2}}\subseteq D_{W_{1}}$, where the set
$D_{W}$ is defined as $D_{W}=\{\rho\geq0, ~\textrm{such that}~
 Tr(W\rho)<0\}$.\\
\textbf{Result-1:} Given two non-decomposable witnesses $W_{1}$
and $W_{2}$, $W_{1}$ is finer than $W_{2}$, if $W_{2}$ can be
written as \cite{12}
\begin{eqnarray}
W_{2}= (1-\lambda)W_{1}+\lambda D \label{finer}
\end{eqnarray}
where $D$ is a decomposable witness operator and
$0\leq\lambda<1$ .\\
\textbf{Result-2:} A witness operator $D$ is decomposable iff
\cite{12}
\begin{eqnarray}
Tr(D\rho)\geq0,~~ \textrm{for PPT entangled state }\rho
\label{decom}
\end{eqnarray}
\section{Revisiting the non-decomposable witness by Lewenstein \emph{et. al.}}
Lewenstein et. al. \cite{9,12} studied the edge states extensively
and introduced a non-decomposable witness exclusively for edge
states $\delta$ which was of the form
\begin{eqnarray}
W^{\delta}=P+Q^{T_{B}}-\varepsilon
I,~~P\geq0,~~Q\geq0,~~0<\varepsilon\leq\varepsilon_{0} \label{lew}
\end{eqnarray}
Since edge states are not of full rank neither are their partial
transpose so $P$ and $Q$ can always be chosen as projectors from
the respective kernels of $\delta$ and $\delta^{T_{B}}$.
$\varepsilon_{0}$ was defined as
\begin{eqnarray}
\varepsilon_{0} = inf_{|e,f\rangle} \langle
e,f|P+Q^{T_{B}}|e,f\rangle
\end{eqnarray}
The above mentioned choices entailed that
\begin{eqnarray}
&&Tr(W^{\delta}\sigma)\geq0~~ \forall ~~ \textrm{separable}~~
\sigma~~ \textrm{and}\nonumber\\&& Tr(W^{\delta}\delta)<0
\end{eqnarray}
When we are willing to detect PPT entangled states which are not
edge states through the witness operator $W^{\delta}$ then in this
situation the task becomes very difficult in choosing the positive
semi-definite operators $P$ and $Q$. This is because of the fact
that the given PPT entangled state $\rho$ (not edge state) or the
state described by its partial transposition can be of full rank.
Therefore the detection of PPT entangled state (excluding edge
states) using $W^{\delta}$ turned out to be a difficult task. So
our focus should be on searching the witness operator which can be
easily constructed and also detects PPT entangled state together
with edge states. We start our search by considering a PPT
entangled state $\rho$. Next we impose two assumptions on $\rho$:\\
\textbf{A1:} The PPT entangled state $\rho$ is not an edge
state.\\
\textbf{A2:} $\rho$ is not of full rank but $\rho^{T_{B}}$ is.\\
With these assumptions, $P$ can be chosen as mentioned earlier
i.e. $P$ can be chosen as a projector on the kernel of
$\rho$.Since there are no vectors in the kernel of
$\rho^{T_{B}}$($\rho^{T_{B}}$ is of full rank), we take $Q$ as a
null operator. These choices of $P$ and $Q$ reduces $W^{\delta}$
to $W^{\rho}$, which is given by
\begin{eqnarray}
W^{\rho}=P-\varepsilon'
I,~~P>0,~~0<\varepsilon'\leq\varepsilon_{1} \label{lew1}
\end{eqnarray}
where
\begin{eqnarray}
\varepsilon_{1} = inf_{|e,f\rangle} \langle e,f|P|e,f\rangle
\end{eqnarray}
Thus, the PPT entangled state $\rho$ which satisfies the above
mentioned assumptions can be detected by the non-decomposable
witness operator $W^{\rho}$.\\
Next our task is to show that $W^{\rho}$ is finer than
$W^{\delta}$. To show this we use the result-1.\\
Therefore the result given in (\ref{finer}) clearly demands that
the witness operator (\ref{lew1}) is finer than its counterpart
(\ref{lew}) because (\ref{lew}) can be written as
\begin{eqnarray}
W^{\delta}=(1-\lambda)W^{\rho}+\lambda D,~~0\leq\lambda<1
\end{eqnarray}
taking $D=Q^{T_{B}}$.\\
Thus, $W^{\rho}$ gives us a more general entanglement witness
which can detect some PPT entangled states along with edge
states, or in other words, $W^{\rho}$ is finer than $W^{\delta}$.\\
\textbf{Illustration:} As an illustration we consider the PPT
entangled state \cite{15}
\begin{eqnarray}
 \rho_{\alpha}=\frac{2}{7}|\psi^{+}\rangle\langle
 \psi^{+}|+\frac{\alpha}{7}\rho_{+}+\frac{5-\alpha}{7}\rho_{-}\label{qutrit}
\end{eqnarray}
where
\begin{eqnarray}
&&\rho_{+}=\frac{1}{3}(|01\rangle\langle01|+|12\rangle\langle12|+|20\rangle\langle20|)\nonumber\\&&
\rho_{-}=\frac{1}{3}(|10\rangle\langle10|+|21\rangle\langle21|+|02\rangle\langle02|)\nonumber\\&&
 |\psi^{+}\rangle=\frac{1}{\sqrt{3}}\sum_{i=0}^{2}|ii\rangle
\end{eqnarray}
The state is PPT entangled for $3<\alpha\leq4$ and edge state for
$\alpha=4$. The rank of $\rho_{\alpha}$ is 7 whereas the rank of
$\rho_{\alpha}^{T_{B}}$ is 9. Now using the prescription described
above for the construction of the witness operator (\ref{lew1}),
we can easily construct the witness operator for the PPT entangled
state $\rho_{\alpha}$ as
\begin{eqnarray}
W^{\rho_{\alpha}}= \left(%
\begin{array}{ccccccccc}
  1-\varepsilon' & 0 & 0 & 0 & -1 & 0 & 0 & 0 & 0 \\
  0 & -\varepsilon' & 0 & 0 & 0 & 0 & 0 & 0 & 0 \\
  0 & 0 & -\varepsilon' & 0 & 0 & 0 & 0 & 0 & 0 \\
  0 & 0 & 0 & -\varepsilon' & 0 & 0 & 0 & 0 & 0 \\
  -1 & 0 & 0 & 0 & 2-\varepsilon' & 0 & 0 & 0 & -1 \\
  0 & 0 & 0 & 0 & 0 & -\varepsilon' & 0 & 0 & 0 \\
  0 & 0 & 0 & 0 & 0 & 0 & -\varepsilon' & 0 & 0 \\
  0 & 0 & 0 & 0 & 0 & 0 & 0 & -\varepsilon' & 0 \\
  0 & 0 & 0 & 0 & -1 & 0 & 0 & 0 & 1-\varepsilon' \\
\end{array}%
\right)
\end{eqnarray}
We observe that
$Tr(W^{\rho_{\alpha}}\rho_{\alpha})=-\varepsilon'<0$.
\section{Construction of the witness and its experimental realization}
 In this section we propose a new non-decomposable witness operator and thereafter
 show that it is indeed a non-decomposable witness operator which detects the edge states.
 Also we study its extension in the multipartite system and further discuss its experimental realization.\\
\textbf{Theorem:} An operator $W$ is a non-decomposable witness
operator for an edge state $\delta$ if it can be expressed in the
form
\begin{eqnarray}
W=Q^{T_{B}}-k(I-P) \label{witop}
\end{eqnarray}
where $P$ is a positive semi-definite operator and $Q$ is a
positive definite operator and $T_{B}$
denotes the partial transposition over the second subsystem.\\
\textbf{Proof:} To prove that $W$ is a non-decomposable witness
operator for an edge state $\delta$, it is sufficient to verify
the two witness inequalities given in (\ref{def.3}) for $W$.\\
\textbf{(i)} We have to show that $ Tr(W\sigma)\geq 0~~~ \forall~~
\textrm{separable state}~ \sigma$.
\begin{eqnarray}
Tr(W\sigma)&&= Tr((Q^{T_{B}}-k(I-P))\sigma) \nonumber\\&& =
Tr(Q\sigma^{T_{B}})-k(1-Tr(P\sigma))~~~~ (\textrm{since}~~
Tr(Q^{T_{B}}\sigma)=Tr(Q\sigma^{T_{B}}))\nonumber\\&& =
((1-Tr(P\sigma)))[\frac{Tr(Q\sigma^{T_{B}})}{(1-Tr(P\sigma))}-k]
\end{eqnarray}
We can always select a value of $k$ from the interval $0<k\leq
k_{0}$ so that $Tr(W\sigma)\geq 0$, where $k_{0}$ is given by
\begin{eqnarray}
k_{0}= \textrm{min} \frac{Tr(Q\sigma^{T_{B}})}{1-Tr(P\sigma)}
\label{k1}
\end{eqnarray}
Here the minimum is taken over all separable states $\sigma$.\\
\textbf{(ii)} Now it remains to be shown that $Tr(W\delta)<0$ for
an edge state $\delta$.\\
Since $\delta$ and $\delta^{T_{B}}$ have some vectors in their
kernel so we get some freedom to choose the operators $P$ and $Q$
as the projectors on $ker(\delta)$ and $ker(\delta^{T_{B}})$
respectively. Therefore, we have $Tr(P\delta)=0$ and
$Tr(Q\delta^{T_{B}})=0$.
\begin{eqnarray}
Tr(W\delta)&&=Tr(Q^{T_{B}}\delta)-kTr((I-P)\delta)\nonumber\\&&
 =Tr(Q\delta^{T_{B}})-k(1-Tr(P\delta))~~~ \nonumber\\&&
 =-k \label{theorem}
\end{eqnarray}
Now using the inequality $0<k\leq k_{0}$ and exploiting equations
(\ref{k1}) and (\ref{theorem}), we find that $Tr(W\delta)<0$.
Hence we are able to prove that the non-decomposable witness
operator
proposed in the theorem detects an edge state.\\
\textbf{Corollary}: The non-decomposable witness can also be
constructed as
\begin{eqnarray}
W'=P-k(I-Q^{T_{B}}),~~0<k\leq k_{0},~~P>0,~~Q\geq0\label{witop1}
\end{eqnarray}
where
\begin{eqnarray}
k_{0}= \textrm{min}_{\sigma}
\frac{Tr(P\sigma)}{1-Tr(Q\sigma^{T_{B}})}
\end{eqnarray}
With similar arguments it can be shown that $W'$ also detects edge
states. Particularly if $Q^{T_{B}}=0$, i.e. if the state described
by the partially transposed density operator has no vectors in its
kernel then witness operator (\ref{witop1}) reduces to
(\ref{lew1}). Hence in this case the witness operator
(\ref{witop1}) detects
not only edge states but also other PPT entangled states.\\
 \textbf{Extension of the witness for edge states in 3
qubits:} Since edge states are also found in tripartite systems so
we extend the prescription of our proposed entanglement witness
operator in 3-qubit systems.\\
For a given tripartite edge state $\delta_{tri} \in B(H_{1}
\otimes H_{2} \otimes H_{3}) $, we define the non-decomposable
witness operator as:
\begin{eqnarray}
W_{tri}=Q^{T_{X}}-k_{0}(I-P),~~~~X=1,2,3 \label{multiwitness}
\end{eqnarray}
$P$=Projector on Ker($\delta_{tri}$) and $Q$= Projector on
Ker($\delta_{tri}^{T_{X}}$),
 where $T_{X}$ denotes the transpose taken with respect to any
one of the subsystems. As before we define
\begin{eqnarray}
k_{0}= \textrm{min}
\frac{Tr(Q^{T_{X}}\sigma)}{Tr((I-P)\sigma)}\label{k0}
\end{eqnarray}
where the minimum is taken over all separable states $\sigma$.
\\If now we take $0< k\leq k_{0}$ and use $W_{tri}=Q^{T_{X}}-k(I-P)$, then we
obtain
\begin{eqnarray}
Tr(W_{tri}\delta_{tri})=-k<0
\end{eqnarray}
For the above choice of $k_{0}$  given in (\ref{k0}),we can
always find some $k$ for which $Tr(W_{tri}\sigma)\geq0$.\\
\textbf{Experimental Realization:} Our task is now to show that
our proposed witness operator can be used in an experimental setup
to detect the edge state in a qutrit system. Since entanglement
witnesses are hermitian operators and every hermitian operators
are observables so they are experimentally realizable quantities.
Thus they provide experimental evidence of entanglement present in
the given system. The quantity to be measured is the expectation
value
\begin{eqnarray}
\langle W \rangle = Tr(W\rho)
\end{eqnarray}
Here we rewrite the witness operator defined in (\ref{witop}) for
a certain edge state in a qutrit system in terms of Gell-Mann
matrices \cite{13} and thereby finding the expectation value of
these physical operators in order to experimentally detect
entanglement.\\ The generalized Gell-Mann matrices are higher
dimensional extensions of the Pauli matrices (for qubits) and are
hermitian and traceless. They form an orthogonal set and
basis. In particular, they can be categorized for qutrits as three different types of traceless matrices : \\
(i) three symmetric Gell-Mann matrices
\begin{eqnarray}
\Lambda^{01}_{s}= \left(%
\begin{array}{ccc}
  0 & 1 & 0 \\
  1 & 0 & 0 \\
  0 & 0 & 0 \\
\end{array}%
\right), &  \Lambda^{02}_{s}=
\left(%
\begin{array}{ccc}
  0 & 0 & 1 \\
  0 & 0 & 0 \\
  1 & 0 & 0 \\
\end{array}%
\right), & \Lambda^{12}_{s}=
\left(%
\begin{array}{ccc}
  0 & 0 & 0 \\
  0 & 0 & 1 \\
  0 & 1 & 0 \\
\end{array}%
\right)
\end{eqnarray}
(ii) three antisymmetric Gell-Mann matrices
\begin{eqnarray}
\Lambda^{01}_{a}= \left(%
\begin{array}{ccc}
  0 & -i & 0 \\
  i & 0 & 0 \\
  0 & 0 & 0 \\
\end{array}%
\right), &  \Lambda^{02}_{a}=
\left(%
\begin{array}{ccc}
  0 & 0 & -i \\
  0 & 0 & 0 \\
  i & 0 & 0 \\
\end{array}%
\right),& \Lambda^{12}_{a}=
\left(%
\begin{array}{ccc}
  0 & 0 & 0 \\
  0 & 0 & -i \\
  0 & i & 0 \\
\end{array}%
\right)
\end{eqnarray}
(iii) two diagonal Gell-Mann matrices
\begin{eqnarray}
\Lambda^{0}= \left(%
\begin{array}{ccc}
  1 & 0 & 0 \\
  0 & -1 & 0 \\
  0 & 0 & 0 \\
\end{array}%
\right), &  \Lambda^{1}=
\left(%
\begin{array}{ccc}
  1/\sqrt{3} & 0 & 0 \\
  0 & 1/\sqrt{3} & 0 \\
  0 & 0 & -2/\sqrt{3} \\
\end{array}%
\right)
\end{eqnarray}
Let us consider a qutrit described by the density operator
(\ref{qutrit}). Our prescribed witness (\ref{witop}) for the state
with $\alpha=4$ is given in matrix form as
\begin{eqnarray}
A= \left(%
\begin{array}{ccccccccc}
  0 & 0 & 0 & 0 & -2 & 0 & 0 & 0 & -k-2 \\
  0 & 1-k & 0 & 0 & 0 & 0 & 0 & 0 & 0 \\
  0 & 0 & 4-k & 0 & 0 & 0 & 0 & 0 & 0 \\
  0 & 0 & 0 & 4-k & 0 & 0 & 0 & 0 & 0 \\
  -2 & 0 & 0 & 0 & 0 & 0 & 0 & 0 & -k-2 \\
  0 & 0 & 0 & 0 & 0 & 1-k & 0 & 0 & 0 \\
  0 & 0 & 0 & 0 & 0 & 0 & 1-k & 0 & 0 \\
  0 & 0 & 0 & 0 & 0 & 0 & 0 & 4-k & 0 \\
  -k-2 & 0 & 0 & 0 & -k-2 & 0 & 0 & 0 & k \\
\end{array}%
\right)
\end{eqnarray}
Writing the witness $A$ in terms of the Gell-Mann matrices and
taking the expectation value we obtain,
\begin{eqnarray}
&&\langle A \rangle=-\langle \Lambda_{s}^{01}\otimes
\Lambda_{s}^{01} \rangle + \langle \Lambda_{a}^{01}\otimes
\Lambda_{a}^{01} \rangle -\frac{k+2}{2}(\langle
\Lambda_{s}^{02}\otimes \Lambda_{s}^{02} \rangle - \langle
\Lambda_{a}^{02}\otimes \Lambda_{a}^{02} \rangle)
\nonumber\\&&-\frac{k+2}{2}(\langle \Lambda_{s}^{12}\otimes
\Lambda_{s}^{12} \rangle - \langle \Lambda_{a}^{12}\otimes
\Lambda_{a}^{12})\rangle
+\frac{2k-5}{4}\langle\Lambda^{0}\otimes\Lambda^{0}\rangle-\frac{9}{4\sqrt{3}}(\langle\Lambda^{0}\otimes\Lambda^{1}\rangle
-\langle\Lambda^{1}\otimes\Lambda^{0}\rangle)\nonumber\\&&
+\frac{22k-45}{36}\langle\Lambda^{1}\otimes\Lambda^{1}\rangle-\frac{k}{9}(\langle\Lambda^{1}\otimes
I \rangle+ \langle I \otimes \Lambda^{1}\rangle)+
\frac{15-5k}{9}\langle I \otimes I\rangle
\end{eqnarray}
Thus for an experimental outcome $\langle A \rangle<0$, the state
is entangled.\\ For qutrits the Gell-Mann matrices can be
expressed in terms of eight physical operators , the observables
$S_{x}, S_{y}, S_{z} ,S_{x}^2 ,S_{y}^2 ,S_{z}^2 ,\{S_{x},S_{y}\},
\{S_{y},S_{z}\}, \{S_{z},S_{x}\} $ of a spin-1 system , where
$\overrightarrow{S}=\{S_{x}, S_{y}, S_{z}\}$ is the spin operator
and $\{S_{i},S_{j}\}= S_{i}S_{j}+S_{j}S_{i}$ (with $i,j=x,y,z$)
denotes the corresponding anticommutator. The representation of
the Gell-Mann matrices in terms of the the spin-1 operators is as
follows \cite{13}:
\begin{eqnarray}
\Lambda_{s}^{01}=\frac{1}{\sqrt{2}\hbar^{2}}(\hbar
S_{x}+\{S_{z},S_{x}\}),&&
\Lambda_{s}^{02}=\frac{1}{\hbar^{2}}(S_{x}^2-S_{y}^2),\nonumber\\
\Lambda_{s}^{12}=\frac{1}{\sqrt{2}\hbar^{2}}(\hbar
S_{x}-\{S_{z},S_{x}\}),&&
\Lambda_{a}^{01}=\frac{1}{\sqrt{2}\hbar^{2}}(\hbar
S_{y}+\{S_{y},S_{z}\}),\nonumber\\
\Lambda_{a}^{02}=\frac{1}{\hbar^{2}}\{S_{x},S_{y}\},&&
\Lambda_{a}^{12}=\frac{1}{\sqrt{2}\hbar^{2}}(\hbar
S_{y}-\{S_{y},S_{z}\}),\nonumber\\
\Lambda^{0}= 2I + \frac{1}{2\hbar^{2}}(\hbar
S_{z}-3S_{x}^2-3S_{y}^2),&&
 \Lambda^{1}= \frac{1}{\sqrt{3}}(-2I+\frac{3}{2\hbar^{2}}(\hbar S_{z}+S_{x}^2+S_{y}^2))
\end{eqnarray}
All eight physical operators can be represented by the following
matrices :
\begin{eqnarray}
S_{x}=\frac{\hbar}{\sqrt{2}}\left(%
\begin{array}{ccc}
  0 & 1 & 0 \\
  1 & 0 & 1 \\
  0 & 1 & 0 \\
\end{array}%
\right), & S_{y}=\frac{\hbar}{\sqrt{2}}\left(%
\begin{array}{ccc}
  0 & -i & 0 \\
  i & 0 & -i \\
  0 & i & 0 \\
\end{array}%
\right), & S_{z}=\hbar\left(%
\begin{array}{ccc}
  1 & 0 & 0 \\
  0 & 0 & 0 \\
  0 & 0 & -1 \\
\end{array}%
\right)\nonumber\\
S_{x}^2= \frac{\hbar^{2}}{2}\left(%
\begin{array}{ccc}
  1 & 0 & 1 \\
  0 & 2 & 0 \\
  1 & 0 & 1 \\
\end{array}%
\right), &
S_{y}^2= \frac{\hbar^{2}}{2}\left(%
\begin{array}{ccc}
  1 & 0 & -1 \\
  0 & 2 & 0 \\
  -1 & 0 & 1 \\
\end{array}%
\right)\nonumber\\
\{S_{x},S_{y}\}= \hbar^{2}\left(%
\begin{array}{ccc}
  0 & 0 & -i \\
  0 & 0 & 0 \\
  i & 0 & 0 \\
\end{array}%
\right), & \{S_{y},S_{z}\}=\frac{\hbar^{2}}{\sqrt{2}}\left(%
\begin{array}{ccc}
  0 & -i & 0 \\
  i & 0 & i \\
  0 & -i & 0 \\
\end{array}%
\right), \nonumber\\
 \{S_{z},S_{x}\}=\frac{\hbar^{2}}{\sqrt{2}}\left(%
\begin{array}{ccc}
  0 & 1 & 0 \\
  1 & 0 & -1 \\
  0 & -1 & 0 \\
\end{array}%
\right)
\end{eqnarray}
Therefore experimental detection of entanglement can also be done
by writing the Gell-Mann matrices in terms of spin-1 operators and
then taking the expectation value.
\section{Detection of a larger set of PPT entangled states by our proposed witness operator}
In this section we study the efficiency of our proposed witness
operator (\ref{witop}). We will show some specific situations in
which our proposed witness operator is finer than its counterpart
in (\ref{lew}).
Thus our operator witness a larger set of PPT entangled states.\\
Now let us recall two witness operators $W^{\delta}$ and $W$ given
in (\ref{lew}) and (\ref{witop}) respectively and investigate the
situation when $W^{\delta}$ detects larger set of PPT entangled
state than $W$ or vice-versa. Also we observe that
\begin{eqnarray}
D_{W}\cap D_{W^{\delta}}\neq \phi \label{intersection}
\end{eqnarray}\\
Equation (\ref{intersection}) depicts the fact that there exist
PPT entangled states which are detected by both $W$ and $W^{\delta}$.\\
\textbf{Case-I:} If the entanglement witness $W$ be finer than
$W^{\delta}$
 then using (\ref{finer}), we can always write
\begin{eqnarray}
&&W^{\delta}= (1-\lambda)W +\lambda D   \nonumber\\&& \Rightarrow
P+Q^{T_{B}}-\varepsilon I = (1-\lambda)(Q^{T_{B}}-k(I-P))+\lambda
D\nonumber\\&& \Rightarrow D = \frac{1-k+\lambda k}{\lambda}P+
Q^{T_{B}} + \frac{k-\varepsilon-\lambda k}{\lambda}I
\label{finer1}
\end{eqnarray}
From (\ref{finer1}) and using the result-2, we get
\begin{eqnarray}
1-k+\lambda k \geq 0, && k-\varepsilon-\lambda k \geq 0
\end{eqnarray}
which gives
\begin{eqnarray}
k\leq \frac{1}{1-\lambda}, && k\geq \frac{\varepsilon}{1-\lambda}
\end{eqnarray}
Thus $W$ is finer than $W^{\delta}$ when $k \in
[\frac{\varepsilon}{1-\lambda}, \frac{1}{1-\lambda}]$.\\
\textbf{Case-II:} If the entanglement witness $W^{\delta}$ be
finer than $W$ then we can proceed in similar way as above and
find that $W^{\delta}$ is finer than $W$ when $k \in [1-\lambda,
\varepsilon-\lambda \varepsilon]$.
\section{Examples}
In this section we explicitly construct our proposed witness
operator for different edge states living in $C^{3} \otimes C^{3}$
and $C^{2} \otimes C^{2} \otimes C^{2}$ and express them in the matrix form.\\
\textbf{Example 1:} We start with the edge state in $C^{3}
\bigotimes C^{3}$ as proposed in \cite{7}. The state and its
partial transpose is :
\begin{eqnarray}
\rho_{a}= \frac{1}{8a+1}
\left(%
\begin{array}{ccccccccc}
  a & 0 & 0 & 0 & a & 0 & 0 & 0 & a \\
  0 & a & 0 & 0 & 0 & 0 & 0 & 0 & 0 \\
  0 & 0 & a & 0 & 0 & 0 & 0 & 0 & 0 \\
  0 & 0 & 0 & a & 0 & 0 & 0 & 0 & 0 \\
  a & 0 & 0 & 0 & a & 0 & 0 & 0 & a \\
  0 & 0 & 0 & 0 & 0 & a & 0 & 0 & 0 \\
  0 & 0 & 0 & 0 & 0 & 0 & \frac{1+a}{2} & 0 &  \frac{\sqrt{1-a^{2}}}{2}\\
  0 & 0 & 0 & 0 & 0 & 0 & 0 & a & 0 \\
  a & 0 & 0 & 0 & a & 0 & \frac{\sqrt{1-a^{2}}}{2} & 0 & \frac{1+a}{2} \\
\end{array}%
\right)
\end{eqnarray}
\begin{eqnarray}
 \rho^{T_{B}}_{a}=\frac{1}{8a+1}
\left(%
\begin{array}{ccccccccc}
  a & 0 & 0 & 0 & 0 & 0 & 0 & 0 & 0 \\
  0 & a & 0 & a & 0 & 0 & 0 & 0 & 0 \\
  0 & 0 & a & 0 & 0 & 0 & a & 0 & 0 \\
  0 & a & 0 & a & 0 & 0 & 0 & 0 & 0 \\
  0 & 0 & 0 & 0 & a & 0 & 0 & 0 & 0 \\
  0 & 0 & 0 & 0 & 0 & a & 0 & a & 0 \\
  0 & 0 & a & 0 & 0 & 0 & \frac{1+a}{2} & 0 &  \frac{\sqrt{1-a^{2}}}{2}\\
  0 & 0 & 0 & 0 & 0 & a & 0 & a & 0 \\
  0 & 0 & 0 & 0 & 0 & 0 & \frac{\sqrt{1-a^{2}}}{2} & 0 & \frac{1+a}{2} \\
\end{array}%
\right)
\end{eqnarray}
where $0<a<1$.\\
The projector on the kernel of $\rho_{a}$ is:
\begin{eqnarray}
P=|00\rangle\langle00|+c|00\rangle\langle20|-|00\rangle\langle22|+c|20\rangle\langle00|+\\\nonumber
  c^{2}|20\rangle\langle20|-c|20\rangle\langle22|-|22\rangle\langle00|-c|22\rangle\langle20|+\\\nonumber
  |22\rangle\langle22|+|11\rangle\langle11|+c|11\rangle\langle20|-|11\rangle\langle22|+\\\nonumber
  +c|20\rangle\langle11|+c^{2}|20\rangle\langle20|-c|20\rangle\langle22|-|22\rangle\langle11|+\\\nonumber
  -c|22\rangle\langle20|+|22\rangle\langle22|
\end{eqnarray}
The partial transpose of the projector on the kernel of
$\rho^{T_{B}}_{a}$ is:
\begin{eqnarray}
Q^{T_{B}}=d^{2}|02\rangle\langle02|-d^{2}|00\rangle\langle22|-d|02\rangle\langle22|-d^{2}|22\rangle\langle00|\\\nonumber
          +d^{2}|20\rangle\langle20|+d|22\rangle\langle20|-d|22\rangle\langle02|+d|20\rangle\langle22|\\\nonumber
          +|22\rangle\langle22|+|12\rangle\langle12|-|11\rangle\langle22|-|22\rangle\langle11|\\\nonumber
          +|21\rangle\langle21|+|01\rangle\langle01|-|00\rangle\langle11|-|11\rangle\langle00|+|10\rangle\langle10|
\end{eqnarray}
where $c=\frac{\sqrt{1-a^{2}}}{1+a}$ and
$d=\frac{\sqrt{1-a^{2}}}{a-1}$. Thus the witness is obtained as :
\begin{eqnarray}
W=\left(%
\begin{array}{ccccccccc}
  0 & 0 & 0 & 0 & -1 & 0 & ck & 0 & -(d^{2}+k) \\
  0 & 1-k & 0 & 0 & 0 & 0 & 0 & 0 & 0 \\
  0 & 0 & d^{2}-k & 0 & 0 & 0 & 0 & 0 & -d \\
  0 & 0 & 0 & 1-k & 0 & 0 & 0 & 0 & 0 \\
  -1 & 0 & 0 & 0 & 0 & 0 & ck & 0 & -1-k \\
  0 & 0 & 0 & 0 & 0 & 1-k & 0 & 0 & 0 \\
  ck & 0 & 0 & 0 & ck & 0 & 2c^{2}k+d^{2}-k & 0 & d-2ck \\
  0 & 0 & 0 & 0 & 0 & 0 & 0 & 1-k & 0 \\
  -(d^{2}+k) & 0 & -d & 0 & -1-k & 0 & d-2ck & 0 & 1+k \\
\end{array}%
\right)\label{witness1}
\end{eqnarray}
Using $W$ as constructed in (\ref{witness1}) we obtain,
\begin{eqnarray}
Tr(W\rho_{a})=-k<0
\end{eqnarray}
\textbf{Example 2:} Next we construct the witness for the edge
state in 3 qubits proposed in \cite{14}. The edge state was
proposed as:
\begin{eqnarray}
\delta_{tri}= \frac{1}{n}
\left(%
\begin{array}{cccccccc}
  1 & 0 & 0 & 0 & 0 & 0 & 0 & 1 \\
  0 & a & 0 & 0 & 0 & 0 & 0 & 0 \\
  0 & 0 & b & 0 & 0 & 0 & 0 & 0 \\
  0 & 0 & 0 & c & 0 & 0 & 0 & 0 \\
  0 & 0 & 0 & 0 & \frac{1}{c} & 0 & 0 & 0 \\
  0 & 0 & 0 & 0 & 0 & \frac{1}{b} & 0 & 0 \\
  0 & 0 & 0 & 0 & 0 & 0 & \frac{1}{a} & 0 \\
  1 & 0 & 0 & 0 & 0 & 0 & 0 & 1 \\
\end{array}%
\right)
\end{eqnarray}
where $n=2+a+b+c+1/a+1/b+1/c$ and the basis is taken in the order
$|000\rangle,|001\rangle,|010\rangle,|011\rangle,\\|100\rangle,
|101\rangle,|110\rangle,|111\rangle$.The partial transpose with
respect to system $C$ is given by:
\begin{eqnarray}
\delta_{tri}^{T_{C}}= \frac{1}{n}
\left(%
\begin{array}{cccccccc}
  1 & 0 & 0 & 0 & 0 & 0 & 0 & 0\\
  0 & a & 0 & 0 & 0 & 0 & 1 & 0 \\
  0 & 0 & b & 0 & 0 & 0 & 0 & 0 \\
  0 & 0 & 0 & c & 0 & 0 & 0 & 0 \\
  0 & 0 & 0 & 0 & \frac{1}{c} & 0 & 0 & 0 \\
  0 & 0 & 0 & 0 & 0 & \frac{1}{b} & 0 & 0 \\
  0 & 1 & 0 & 0 & 0 & 0 & \frac{1}{a} & 0 \\
  0 & 0 & 0 & 0 & 0 & 0 & 0 & 1 \\
\end{array}%
\right)
\end{eqnarray}
The vector in the kernel of $\delta_{tri}$ is
$|000\rangle-|111\rangle$ and the vector in the kernel of
$\delta_{tri}^{T_{C}}$ is $|001\rangle-a|110\rangle$. With these
vectors the witness (\ref{multiwitness}) is obtained as :
\begin{eqnarray}
W_{tri}=
\left(%
\begin{array}{cccccccc}
  0 & 0 & 0 & 0 & 0 & 0 & 0 & -k-a\\
  0 & 1-k & 0 & 0 & 0 & 0 & 0 & 0 \\
  0 & 0 & -k & 0 & 0 & 0 & 0 & 0 \\
  0 & 0 & 0 & -k & 0 & 0 & 0 & 0 \\
  0 & 0 & 0 & 0 & -k & 0 & 0 & 0 \\
  0 & 0 & 0 & 0 & 0 & -k & 0 & 0 \\
  0 & 0 & 0 & 0 & 0 & 0 & a^{2}-k & 0 \\
  -k-a & 0 & 0 & 0 & 0 & 0 & 0 & 0 \\
\end{array}%
\right)
\end{eqnarray}
which gives,
\begin{eqnarray}
Tr(W_{tri}\delta_{tri})=-k<0
\end{eqnarray}

\section{Conclusion}
To summarize, we have constructed a non-decomposable witness
operator which gives a negative expectation value on edge
states,thereby detecting them. Our proposed witness operator is
interesting in the sense that it sometime detects larger set of
PPT entangled state than the non-decomposable witness operator
given by Lewenstein et.al.\cite{12}. In technical terms we have
showed that our proposed witness operator is finer than the
Lewenstein et.al. operator in some situation and found that in
some cases the fact is reverse. Also we have discussed its
experimental relevance to substantiate the worth of the witness,
which to our knowledge can help us to detect PPT entangled states
experimentally.
\section{Acknowledgement}
NG acknowledges his mother for her eternal love and blessings.


\begin{thebibliography}{99}
\bibitem{1} A. Einstein, B. Podolsky and N. Rosen, Phys. Rev.
\textbf{47}, 777 (1935).
\bibitem{2} E. Schrodinger, Naturewissenschaften \textbf{23}, 807 (1935).
\bibitem{3} A. Barenco, D. Deutsch, A. Ekert and R. Jozsa, Phys. Rev.
Lett. \textbf{74}, 4083 (1995).
\bibitem{4} C. H. Bennett, G. Brassard, C. Crepeau, R. Jozsa, A. Peres and
W. K. Wootters, Phys. Rev. Lett. \textbf{70}, 1895 (1993).
\bibitem{5} A. Peres, Phys. Rev. Lett. \textbf{77}, 1413 (1996).
\bibitem{6} M. Horodecki, P. Horodecki and R. Horodecki, Phys. Lett. A
\textbf{223}, 1 (1996).
\bibitem{7} P. Horodecki ,Phys. Lett. A \textbf{232}, 333 (1997).
\bibitem{8} C. H. Bennett, D. P. DiVincenzo, T. Mor, P. W. Shor, J. A. Smolin
and B. M. Terhal, Phys. Rev. Lett. \textbf{82}, 5385 (1999).
\bibitem{9} M. Lewenstein, B. Krauss, P. Horodecki and
J. I. Cirac, arxiv:quant-ph/0005112v1 (2000).
\bibitem{10} B. M. Terhal, arxiv:quant-ph/9810091.
\bibitem{11} S. L. Woronowicz, Rep. Math. Phys. \textbf{10}, 165
(1976).
\bibitem{12} M. Lewenstein, B. Krauss, J. I. Cirac and P. Horodecki, Phys. Rev. A \textbf{62}, 052310
(2000).
\bibitem{13} R. A. Bertlmann and P. Krammer, J. Phys. A: Math. Theor \textbf{41}, 235303 (2008).
\bibitem{14} A. Acin, D. Bruss, M. Lewenstein and A. Sanpera, Phys. Rev. Lett. \textbf{87}, 040401 (2001).
\bibitem{15} P. Horodecki, M. Horodecki, R. Horodecki, Phys. Rev. Lett. \textbf{82},
1056 (1999).
\end{thebibliography}
\end{document}